\documentclass[12pt,preprint]{aastex}
\usepackage{emulateapj5,epsf}

\shorttitle{Globular Cluster MBH}
\shortauthors{Kawakatu and Umemura}

\begin{document}

\title{Formation of Massive Black Holes in Globular Clusters}


\author{Nozomu Kawakatu\altaffilmark{1}}
\affil{International School for Advanced Studies (SISSA/ISAS), Via Beirut 2-4, 34014 Trieste, Italy}

\and 

\author{Masayuki Umemura\altaffilmark{2}}
\affil{Center for Computational Sciences, University of
  Tsukuba, Ten-nodai, 1-1-1 Tsukuba, Ibaraki, 305-8577, Japan}

\altaffiltext{1}{kawakatu@sissa.it}
\altaffiltext{2}{umemura@css.tsukuba.ac.jp}



\begin{abstract}

The hydrodynamic formation of massive black holes (BHs)
in globular clusters is considered.
In particular, we examine the possibility of the BH formation
induced by the radiation drag that is exerted on the interstellar 
matter by stellar radiation in globular clusters. 
The radiation drag extracts angular momentum from interstellar gas 
and thus allows the gas to accrete on to the cluster center.
By incorporating the realistic chemical evolution of 
globular clusters, we scrutinize the efficiency of the radiation drag 
to assess the total accreted mass. 
As a result, we find that 
if a globular cluster is more massive than
$\approx 6\times 10^{6}M_{\odot}$ 
in the present-day stellar component,
a massive BH with $>260M_{\odot}$ can form within it. 
But, the BH-to-cluster mass ratio 
is considerably smaller than the BH-to-bulge mass ratio ($\approx 10^{-3}$) 
found in galactic bulges. 
The results are not sensitive to the assumed stellar initial mass function and
star formation rate in the cluster, 
as long as the resultant color-magnitude relation, metallicity, 
and mass-to-luminosity ratio satisfy those observed in globular clusters. 
Hence, the putative linear relation between BH mass and bulge mass  
($M_{\rm BH}-M_{\rm bulge}$ relation) cannot be extrapolated 
to globular cluster systems. 
In the present regime, we discuss the BH formation
in M15, G1, $\omega$ Cen, the M33 nucleus, 
and the compact X-ray sources in M82. 
Finally, we argue observational indications of the formation process 
of massive BHs in globular clusters. 
We find that the final phase of BH growth due to the radiation drag 
can be observed as ultraluminous X-ray sources (ULXs) 
with $\sim 10^{41}{\rm erg}\,{\rm s}^{-1}$. 

\end{abstract}

\section{Introduction}
The recent compilation of the kinematical data on galactic centers
has revealed that
a central ``massive dark object ''(MDO), which
is a candidate for suppermassive black hole (BH),
does correlate with the mass of a galactic bulge; the BH-to-bulge mass ratio 
is $\approx 0.002$ as a median value (e.g., Kormendy \& Richstone 1995).
This correlation suggests that the formation of a supermassive BH is 
physically connected with the formation of a galactic bulge.

If the BH-to-bulge relation can be extrapolated to small spheroidal
systems like globular clusters, massive BHs with $10^{3-4}M_\odot$
may inhabit globular clusters.
To date, some candidates of massive BHs in globular clusters have
been reported; the estimated mass of a possible BH in M15 is
$M_{\rm BH} =(1.7^{+2.7}_{-1.7}) \times 10^{3}M_{\odot}$ 
 (Gerssen et al. 2003), or
$M_{\rm BH}\approx 2\times 10^{4}M_{\odot}$ 
in G1 (Gebhardt, Rich \& Ho 2002).
On the contrary, recent sophisticated numerical simulations (Baumgardt et al. 2003a, b)
show that a massive BH with $> 500-1000M_{\odot}$
is not mandatory to account for the observational data in M15, 
and the luminosity profile of G1 can be well fitted without a massive BH. 
On the other hand,
M33 nucleus does not appear to possess a massive BH 
with an upper limit of $\approx 10^{3}M_{\odot}$ 
(Gebhardt et al. 2001; Merritt, Ferrarese \& Joseph 2001). 
In addition, 
latest radio observations have suggested that 
the BH mass in $\omega$ Cen should be less than about 
$100 M_{\odot}$ for the spherical Bondi-Hoyle accretion rate, but
the data are marginally consistent with a BH of 
about $1000 M_{\odot}$ for more plausible accretion rate
 (Maccarone et al. 2005).
Thus, the existence of massive BHs in globular clusters is under debate.

Some theoretical models have been proposed for the formation 
of massive BHs. One of them is the merger of multiple 
small BHs or stars 
(e.g., Lee 1987; Quinlan \& Shapiro 1990).
However, this scenario predicts that, 
even in a very compact star cluster
like a globular cluster,
the core-collapse is halted by the binary heating 
before the stellar density becomes high enough for stars 
to merge in a runaway fashion (Hut et al. 1992). 
However, it is pointed out recently that the mass difference of 
constituents plays an important role.
Stars of different masses are not always able to reach the energy equipartition. 
This causes the heaviest stars to undergo core collapse on a time scale 
that is much shorter than the core collapse time for the cluster as a whole. 
Portegies Zwart \& McMillan (2002) show by $N$-body simulations that 
a runaway merger among these massive stars leads to the formation of a massive BH.
On the other hand, Miller \& Hamilton (2002) examine a model where a single BH 
($> 50M_{\odot}$) located at the center of the cluster grows in mass 
through merging with stellar mass BHs.

Another possibility is the hydrodynamic formation of massive BHs
through supermassive stars. 
Recently, as a potential hydrodynamic mechanism in a bulge, 
Umemura (2001) has considered the effects of radiation drag,
which is equivalent to a well-known Poynting-Robertson effect 
in the solar system. 
The radiation drag extracts angular momentum from interstellar gas 
and thus allows the gas to accrete on to the center. 
This is a parallel model of the formation of massive BHs by the 
Compton drag in the early universe (Umemura, Loeb, \& Turner 1997). 
The mass accretion by the radiation drag has been explored in detail
(Umemura, Fukue, \& Mineshige 1997; Fukue, Umemura, \& Mineshige 1997).
The angular momentum loss rate by the radiation drag is given by
$d \ln J/dt \simeq -\chi E/c$, 
where $J$ is the total angular momentum of gaseous component, $E$ is the radiation energy density, and $\chi$ is the mass extinction coefficient. 
Therefore, in an optically-thin regime, $d \ln J/dt \simeq -(\tau L_{*}/c^{2}M_{\rm gas})$, where $\tau$ is the total optical depth of the system, $L_{*}$ is the total luminosity of the spheroidal system and $M_{\rm gas}$ is the total mass of gas. In an optically thick regime, the 
radiation drag efficiency is saturated due to the conservation of photon number (Tsuribe \& Umemura 1997). Thus, an expression of the angular momentum loss rate which includes both regimes is given by 
$d \ln J/dt \simeq -(L_{*}/c^{2}M_{\rm g})(1-e^{-\tau})$.
Then, the mass accretion rate is estimated to be 
$\dot{M}=-M_{\rm g}d \ln J/dt = L_{*}/c^{2}(1-e^{-\tau})$. 
In an optically-thick regime, this gives simply 
$\dot{M}= L_{*}/c^{2}$ (Umemura 2001). 
Then, the total accreted mass on to the MDO, $M_{\rm MDO}$, is maximally 
$M_{\rm MDO} \simeq \int {L_{*}}/{c^{2}}dt$.
Although this is a simple estimation in one-zone interstellar medium (ISM), the ISM is observed to be highly inhomogeneous 
in an active star-forming galaxy 
(Sanders et al. 1988; Gordon, Calzetti \& Witt 1997).
In such an inhomogeneous ISM,
optically-thin surface layers of optically-thick clumpy clouds
are stripped by the radiation drag, 
and the stripped gas loses angular momentum, eventually
accreting on to the center (Sato et al. 2004).
Kawakatu \& Umemura (2002) have shown that the inhomogeneity of 
ISM plays an important role 
for the radiation drag to attain the maximal efficiency.
Then, the final mass of MDO is proportional to 
the total radiation energy from stars.
Taking the realistic chemical evolution into consideration, 
the radiation drag model predicts a mass ratio as
$M_{\rm BH}/M_{\rm bulge}\simeq 0.001$ in a galactic bulge
(Kawakatu, Umemura, \& Mori 2003).
The theoretical upper limit of BH-to-bulge mass ratio 
is determined by the energy conversion efficiency 
of nuclear fusion from hydrogen to helium, i.e., 0.007 (Umemura 2001).
Here, the question is whether the radiation drag mechanism 
can work also in small spheroidal systems like globular clusters.

In this paper, we examine the possibility of the radiation drag-induced
formation of massive BHs in globular clusters. 
Here, we assume that the formation of a globular cluster begins
with the coeval starburst within it.
The outline of the physical processes we consider is shown
in Figure 1: 
(I) First, the energy input from type II supernovae (SNe) 
drives the outflow of ISM from the system.
Therefore, the system becomes
optically-thin in an early evolutionary phase.
(II) Afterward, intermediate mass ($2-8M_{\odot}$) stars 
shed the gas envelope into the interstellar space, 
accumulating the ISM, and thus
the optical depth of the system increases. 
In this phase, the radiation drag can work effectively
and the mass accretion on to the center is induced.
(III) Finally, type Ia SNe, with type II SNe in newly formed stars,
expel the ISM from the system again. 
Resultantly, the efficiency of the radiation drag descends abruptly.
Through these physical processes, it is expected that 
the BH formation in a globular cluster is strongly related to
the star formation history. 



In this paper, we attempt to elucidate whether
the formation of massive BHs can be prompted by a radiation
hydrodynamic mechanism in globular clusters.
Also, based on the present model, we reconsider
the formation of massive BHs in M15, G1, $\omega$ Cen, 
the M33 nucleus, and the compact X-ray sources in M82. 
The paper is organized as follows.
In Section 2, we build up the model for the chemical evolution and
the radiation-hydrodynamic process.
In Section 3, we investigate the relation between the star formation history 
and the final BH mass in a cluster. 
Then, we derive the condition on which 
globular clusters can possess massive BHs.
In Section 4, we compare our predictions with the observational results 
for M15, G1, $\omega$ Cen, the M33 nucleus, and the compact X-ray sources in M82,
and discuss whether massive BHs can form through
radiation-hydrodynamic process in these globular clusters. 
In Section 5, based on the present regime, we discuss observational 
indications of the formation process of massive BHs in globular clusters. 
Scetion 6 is devoted to the conclusions.

\vspace{5mm}
\epsfxsize=8cm 
\epsfbox{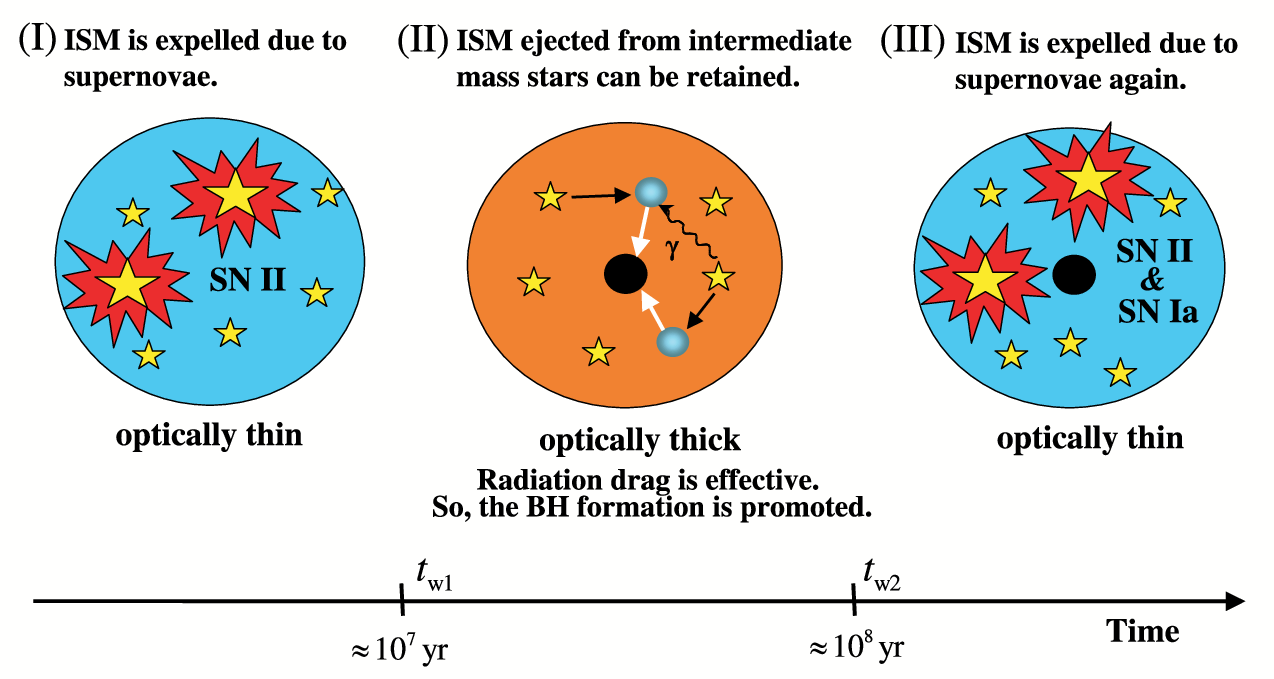}
\figcaption
{
The outline of the present scenario for the formation of a massive BH 
in a globular cluster.
}

\section{Models}
\subsection{Chemical evolution of globular cluster}
We suppose a two-component system that consists of a spheroidal 
stellar cluster and dusty interstellar matter within it.
First, we construct the model for the chemical evolution of globular clusters, by using an evolutionary spectral synthesis 
code 'PEGASE' (Fioc \& Rocca-Volmerange 1997). 
At the initial epoch, we set a spherical system composed of gas
with mass given in the range of $M_{0}=10^{5-9}M_{\odot}$. 
The radius, $r_{\rm s}$, is determined to follow a mass-radius 
relation, $(r_{\rm s}/10{\rm pc})=(M_{0}/10^{6}M_{\odot})^{1/2} $,
found by Chiosi \& Carraro (2002). 
It is assumed that the star formation starts with 
an overall starburst with a duration of $\approx 10^{7}$yr, which
reproduces the color-magnitude relation of globular clusters 
at the present epoch (Yoshii \& Arimoto 1987). 
We employ a stellar initial mass function (IMF) in the form of
$\phi = dn/d\log m_{*}=A(m_{*}/M_{\odot})^{-\alpha}$, where
$m_{*}$ is the stellar mass. 
The lower mass in the IMF is assumed to be $m_{\rm l}=0.1M_{\odot}$ 
and the upper mass to be $m_{\rm u}=60M_{\odot}$.
The index $\alpha$ is changed from 0.2 to 2.0 in order to see
the effects of the slope of IMF on the final results. 
Also, stars with masses greater than $8M_{\odot}$
are postulated to undergo type II supernova explosions. 
As for the type Ia supernova, an evolutionary spectral syntheseis 
code 'PEGASE' adopts the single-degenarate scenario (Nomoto et al. 1994) 
for the type Ia supernova progenitor, where a white dwarf (WD) in a 
close binary undergoes a thermonuclear explosion when the companion star evolves off the main sequence and transfers a large enough amount of mass to the WD. 
\footnote{In present-day galaxies, it is well known that more than half of main-sequence stars are observed in binaries, while it is not observationally clear
if Population III stars are able to form in binary systems. 
But, Saigo, Matumoto \& Umemura (2004) have recently shown by numerical
simulations that a significant fraction of Population III stars are expected to form in binary systems, and thus type Ia supernovae would occur even in the metal-deficient environments.
}
From the evolution model of progenitor stars, Hachisu et al. (1996, 1999) obtained the mass range of companion stars of the WDs that become 
type Ia supernovae as $[0.9M_{\odot}, 3.5M_{\odot}]$.
The star formation rate (SFR) per unit mass , $C(t)$,
is assumed to be in proportion to the gaseous mass fraction,
$f_{\rm g}(t) \equiv M_{\rm g}(t)/M_{\rm 0}$,
where $M_{\rm g}(t)$ is the total gas mass at time $t$.
The SFR is given by
\begin{equation}
C(t)=\left\{
     \begin{array}{@{\,}ll}
     \beta f_{\rm g} & t < t_{\rm w1}, \\
     kf_{\rm g} & (t_{\rm w1}\leq t < t_{\rm w2}),\\
     0  & (t \geq t_{\rm w2}). \\
    \end{array}
    \right.
\end{equation}
The epoch $t_{\rm w1}$ is determined 
by the age of the minimal star ($8M_{\odot}$) that undergoes 
a type II supernova, that is, $ t_{\rm w1} =6\times 10^{7}$yr.
Before $t_{\rm w1}$, the interstellar gas is assumed to 
blow out from the system due to the energy input from successive
type II SNe.
$t_{\rm w2}$ is the second blowout epoch
when the interstellar gas is expelled by type Ia SNe as well as
type II SNe in newly formed stars.
The second blowout epoch is estimated by the epoch when
the kinetic energy ($E_{\rm kin}$) equals the binding energy ($E_{\rm bin}$).
The kinetic energy is evaluated by 
$E_{\rm kin}=\epsilon 10^{51}N_{\rm SN}{\rm erg}$, where
$N_{\rm SN}$ is the number of SNe, which is 
calculated by 'PEGASE'.
$\epsilon$ is the efficiency at which the total SN energy is converted
to the kinetic energy of blast wave. The Sedov solution gives
$\epsilon=0.28$ for the adiabatic exponent of $\gamma =5/3$.
After $t_{\rm w2}$, 
the system is assumed to be ISM-deficient and thus optically thin.
The coefficient $\beta$ is assumed to be 
$\beta \approx (10^{7}{\rm yr})^{-1}$(Yoshii \& Arimoto 1987), 
and $k$ is changed in order to see the dependence on the SFR.

As for the stellar component, we assume star clusters with a specific 
stellar initial mass function (see above). Stars are distributed uniformly inside the spheroidal system. 
Actually, $N_{*}(=300)$ small star clusters are distributed randomly. 
In addtion, the angular momentum is smeared, assuming rigid rotation, 
according to the spin parameter of $\lambda=(J_{\rm T}|E_{\rm T}|^{1/2})/(GM_{\rm T}^{5/2})=0.05$, where $J_{\rm T}$, $E_{\rm T}$, and $M_{\rm T}$ are respectively the total angular momentum, energy, and mass (Barnes \& Efstathiou 1987; Heavens \& Peacock 1988; Steinmetz \& Bartelmann 1995).

As for ISM, we consider clumpy matter, since the ISM is observed to be highly inhomogeneous in active star-forming regions (Sanders et al. 1988; Gordon, Calzetti, \& Witt 1997). Here, $N_{\rm c}(=10^{4})$ identical clouds are distributed randomly. (It is noted that simulations with a three times larger number of clouds did not lead to any fundamental difference in the final BH mass, although at least
$10^{4}$ clouds are necessary to treat the radiative transfer effect properly in clumpy ISM.) The distribution and motion of ISM is the same as those of the stellar component. 
We assume that the cloud covering factor is unity according to the previous analysis (Kawakatu \& Umemura 2002). The internal density in a cloud is assumed to be uniform, and the optical depth is determined by a dust-to-gas ratio that is calculated by 'PEGASE'. Thus, the total optical depth of globular cluster, $\tau_{\rm T}$ , is a function of time. Also, the optical depth of a cloud and therefore the overall optical depth of globular cluster depend on the cloud size $r_{\rm c}$.
Here, $r_{\rm c}=0.01r_{\rm s}$ is assumed as a fiducial case. 
But, we have confirmed that no essential difference in the final BH mass is found by changing $r_{\rm c}$ so as to enhance $\tau_{\rm T}$ by an order.

\subsection{Mass accretion due to radation drag}
Next, we model the formation of a BH in a globular cluster, based on the 
radiation drag-driven mass accretion.
The radiation drag, which drives the mass accretion, originates in the relativistic effect in absorption and subsequent re-emission of the radiation. This effect is naturally involved in relativistic radiation hydrodynamic equations (Umemura, Fukue, \& Mineshige 1997; Fukue, Umemura, \& Mineshige 1997). 
The angular momentum transfer in radiation hydrodynamics is 
given by the azimuthal equation of motion in cylindrical coordinates, 
\begin{equation}
\frac{1}{r}\frac{d(rv_{\rm \phi})}{dt}=\frac{\chi_{\rm d}}{c}[F^{\rm \phi}-
(E+P^{\rm \phi \phi})v_{\rm \phi}], \label{ldot}
\end{equation} 
where $E$ is the radiation energy density, $F^{\rm \phi}$ 
is the radiation flux, $P^{\rm \phi\phi}$ is the radiation stress tensor,
and $\chi_{\rm d}$ is the mass extinction coefficient which is given 
by $\chi_{\rm d}=n_{\rm d}\sigma_{\rm d}/\rho_{gas}$ with the number density of dust grains $n_{\rm d}$, the dust cross-section $\sigma_{\rm d}$ and the gas density $\rho_{\rm gas}$.
By solving radiative transfer with including dust opacity, 
we evaluate the radiative quantities, $E$, $F^{\rm \phi}$, and $P^{\rm \phi\phi}$, and thereby obtain
the total angular momentum loss rate.
Then, we can estimate the total mass of the dusty ISM accreted on to a central massive drak object, $M_{\rm MDO}$. 
By using the relation, $\dot{M}_{\rm gas}/M_{\rm gas}=-\dot{J}/J$, 
where $J$ and $M_{\rm gas}$ are the total angular momentum and gas of ISM. Then,  
the MDO mass is assessed as     
\begin{equation}
M_{\rm MDO}(t)=\int_{0}^{t}\dot{M}_{\rm gas}dt 
=-\int_{0}^{t}M_{\rm gas}\frac{\dot{J}}{J}dt.
\end{equation}
In the optically-thick regime of the radiation drag,
$M_{\rm gas}\dot{J}/{J}=\int^{\infty}_{0}
[L_{\rm star, \nu} (t)/c^{2}] d\nu$, where $L_{{\rm star},\nu}$ is
the total luminosity of globular cluster at frequency $\nu$.
The radiation drag efficiency depends on the optical depth $\tau$ 
in proportion to $(1-e^{-\tau})$ (Umemura 2001). 
Thus, the total mass of MDO can be expressed by 
\begin{equation}
M_{\rm MDO}(t)=\eta_{\rm drag}\int^{t}_{0}\int^{\infty}_{0}
\frac{L_{{\rm star},\nu}(t)}
{c^2}\left(1-e^{-\tau_{\nu}(t)}\right) d\nu dt,
\end{equation}
where $\tau_{\nu}$ is
the optical depth of the globular cluster measured from the center.
Here, we estimate the evolution of $\tau_{\nu}$ 
by using 'PEGASE'.
The efficiency $\eta_{\rm drag}$ is found to be maximally 0.34 
(Kawakatu $\&$ Umemura 2002).
After $t_{\rm w2}$, the radiation drag is inefficient, 
because the system becomes ISM-deficient and optically thin.
In this paper, at $t < t_{\rm w1}$ and $t > t_{\rm w2}$,
$\tau_{\nu}$ is assumed to drop abruptly to $\tau_{\nu}\ll 1$.
Hence, the final mass of an MDO is given by
\begin{equation}
M_{\rm MDO}=\eta_{\rm drag}\int^{t_{\rm w2}}_{t_{\rm w1}}\int^{\infty}_{0}
\frac{L_{{\rm star},\nu}(t)}{c^2}\left(1-e^{-\tau_{\nu}(t)}\right) d\nu dt.
\label{m_mdo}
\end{equation}

\subsection{The fate of the MDOs}
In the previous section, we estimated the MDO mass in the 
context of the radiation drag-induced mass accretion. 
However, the MDO itself does not evolve into the massive 
BH directly, because the radiation drag is not likely to remove 
the angular momentum thoroughly (Sato et al. 2004). 
Hence, some residual angular momentum will terminate 
the radial contraction of the accreted gas. 
Thus, the MDO is likely to be a compact rotating disk.
We should consider the further collapse of the MDO
through other physical mechanisms. In the MDO, the viscosity is expected
to work effectively, because the timescale for viscous accretion is
shortened by the radiation drag (Mineshige, Tsuribe, \& Umemura 1998). 
Thus, the MDO would be a massive self-gravitating viscous disk. 
For a massive self-gravitating viscous disk,
some self-similar solutions are known to give an inside-out disk collapse 
in a flat temperature distribution of a disk (Mineshige \& Umemura 1996, 1997; Tsuribe 1999). A flat temperature profile is plausible as long as the viscous heating is balanced with the radiative cooling in the geometrically thin disk (see Mineshige \& Umemura 1996 for details). 
In fact, Tsuribe (1999) derived a series of self-similar solutions for 
rotating isothermal disk, taking into account the growth of the central point mass, and then provided a convenient formula for the inside-out mass accretion rate for a variety of outer flows, 
\begin{equation}
\dot{M}_{\alpha}=\frac{3\alpha c_{\rm s}^{3}}{QG},
\end{equation}
where $\alpha$ is the viscous parameter, $c_{\rm s}$ is the sound speed and $Q$ is theToomre's Q which is $\kappa c_{\rm s}/\pi G\Sigma$ for
the epicycle frequency $\kappa$ and the surface density $\Sigma$. 
Tsuribe (1999) has found that an accretion with $Q\approx 2$ 
is stable for a wide range of $\alpha$.
The critical accretion rate is given by 
\begin{equation}
\dot{M}_{\alpha}\simeq \frac{c_{\rm
 s}^{3}}{G}=0.24M_{\odot}{\rm yr}^{-1}
\left(\frac{T_{\rm disk}}{10^{4}K}\right)^{3/2}.
\end{equation}
It must be noted that the viscous accretion rate is
sensitively dependent on the disk temperature. 
Under an intense starburst, the disk is exposed to the
strong ultraviolet radiation.
Then, the disk could be heated up to $10^{4}{\rm K}$,
although the detail is dependent on the dust extinction
and radiative cooling. Here, we employ $10^{4}{\rm K}$ as
the disk temperature. 
Through this inside-out collapse, almost all the MDO can accrete on to 
a centeral core.
Finally, the core is likely to be a rigidly-rotating very massive star (VMS), because the angular momentum transfer via viscosity
works to smear out any differential rotation in a self-gravitating system.
Such VMSs pessess large helium cores that reach carbon ignition. 
The fate of VMSs sensitively depends on the initial mass
(Bond, Arnett, \& Carr 1984; Heger \& Woosley 2002).
A VMS below $260M_{\odot}$ (in the range $140-260M_{\odot}$)
results in a pair-instability supernova, leaving no remnant,
because the core of such a star undergoes
the electron-positron pair creation instability after the helium burning
(Barkat, Rakavy, \& Sack 1967; Bond, Amett, \& Carr 1984; Kippenhahn \& 
Weigert 1990; Flyer, Woosely, \& Herger 2001). 
Above $\sim 260M_{\odot}$, a VMS collapses directly into a BH 
because the nuclear burning cannot halt the core collapse 
(Fryer et al. 2001; Heger et al. 2003). 
There must be the effect of rotation, because the progenitor stars are 
likely to be rotating (Bond et al. 1984; Fryer et al. 2001).
Recently, Shibata (2004) finds that a rigidly-rotating VMS
with several $100 M_\odot$ can be unstable in general relativity 
for softer equations of state, eventually forming a BH. 

Here, we should pay attention to the metallicity dependence of
the VMS formation. In the solar abundance, any stars more massive than
$100 M_\odot$ are not expected to form (Fryer et al. 2001; Heger et al. 2003).
On the other hand, VMSs can preferentially form in metal-free gas 
(e.g Nakamura \& Umemura 2001, and references therein). 
The problem is the critical metallicity, under which
VMSs can form. Recently, Omukai \& Palla (2003) have explored 
this problem in detail, and found the critical metallicity 
to be around $0.01Z_{\odot}$, under which dust opacity is not 
effective to hinder the mass accretion onto a protostar. 
Hence, in the light of the critical metallicity, 
VMSs are expected to form in globular clusters. 
Taking such theoretical advances into account,
we assume $260M_{\odot}$ as the minimum mass for the formation 
of massive BHs. 
If $M_{\rm MDO}$ exceeds this threshold 
mass ($260M_{\odot}$), a massive BH whose mass is equal to a 
MDO mass would form in a globular cluster. 
However, it is not clear enough
whether the self-similar solutions employed in this paper 
are feasible in realistic situations for the MDO formed via the radiation drag.
The problem should be explored with sophisticated numerical simulations,
which are left for the future study.
In this paper, we regard a BH mass estimated here 
as a maximal mass of BH within a globular cluster. 

\section{Results}
By changing physical parameters $\alpha$ and $k$, which control
the star formation, we investigate the relation
between the formation of massive BHs and the star formation history.
First, we examine the dependence on the stellar IMF
and then on the star formation rate. 
Finally, we derive the relation between
the mass of globular cluster and the BH mass.

\subsection{Dependence on Initial Mass Function}

Here, we investigate the effects of the IMF slope
($\alpha$) on the final BH mass, with
settling $M_{0}=10^{8}M_{\odot}$ and $k=8.6 {\rm Gyr}^{-1}$.
In Figure 2, the resultant mass of BH ($M_{\rm BH}$) is 
shown for the range of $\alpha=0.2-2.0$.
The details of the results are summarized in Table 1.
As seen in Figure 2, we find that
the mass of central massive objects
exceeds the threshold mass ($260 M_\odot$) 
for the BH formation in the range of $\alpha=0.2-2.0$.
Also, the BH mass shows a peak around $\alpha \simeq 1$.
The reason can be understood as follows.
First, let us remind that the final BH mass is determined 
by the optical depth of the system and 
the duration between $t_{\rm w1}$ and $t_{\rm w2}$,
as shown in equation (\ref{m_mdo}).
In the present scenario, the optical depth 
is controlled by the mass loss from stars with $<8M_{\odot}$.
If $\alpha$ is smaller, a larger number of massive stars form and
therefore the optical depth can be increased in a shorter timescale 
by the mass loss from intermediate mass stars.
But, simultaneously an increasing number of SN explosions 
shorten the second blowout time $t_{\rm w2}$. 
Table 1 shows that $t_{\rm w2}$
reduces vastly with decreasing $\alpha$, 
so that $M_{\rm BH}$ lessens.
On the other hand, if $\alpha$ is larger, 
the fraction of massive stars decreases. Resultantly, 
fewer SN explosions lengthen $t_{\rm w2}$. But, 
the optical depth cannot augment owing to a reduced number
of intermediate mass stars. As a consequence,
$M_{\rm BH}$ lessens with increasing $\alpha$. 
By these two effects, $\alpha \simeq 1$ leads to the
maximum efficiency of the radiation drag.

The change of $\alpha$ brings a different mass-to-luminosity ratio 
and metallicity of a globular cluster.
The shaded region satisfies the observed mass-to-luminosity ratio
of globular clusters, 1.0-3.0 in solar units 
(e.g., Mandushev et al. 1991; Djorgovski et al. 2003), and also
the metallicity, $\approx 0.01-0.1Z_{\odot}$ 
(Lee 1990; Chaboyer, Demarque \& Sarajedini 1996),
where $Z_{\odot}$ is the solar metallicity.
If the slope of IMF is in the range of $\alpha=0.7-1.7$, 
the properties of globular clusters are consistent 
with the observations.
In this range of $\alpha$, 
massive BHs with $\approx 10^{3}M_{\odot}$ are expected to form.

In this calculation, we have assumed $m_{\rm l}=0.1M_{\odot}$ 
and $m_{\rm u}=60M_{\odot}$.
To see the effect of a different choice of the lower and upper 
mass of the IMF, we also investigated the case of 
$(m_{\rm l}, m_{\rm u})$=$(0.5M_{\odot}, 60M_{\odot})$, 
$(0.1M_{\odot}, 30M_{\odot})$, and
$(0.1M_{\odot}, 120M_{\odot})$, all of which reproduce
the observed mass-luminosity ratio. 
As a result, we found that the final BH mass is not strongly
dependent on $m_{\rm l}$ and $m_{\rm u}$.
To conclude, as long as the mass-to-luminosity ratio 
and metallicity satisfy those observed, the final BH mass
is not sensitive to the stellar initial mass function.

\vspace{5mm}
\epsfxsize=8cm 
\epsfbox{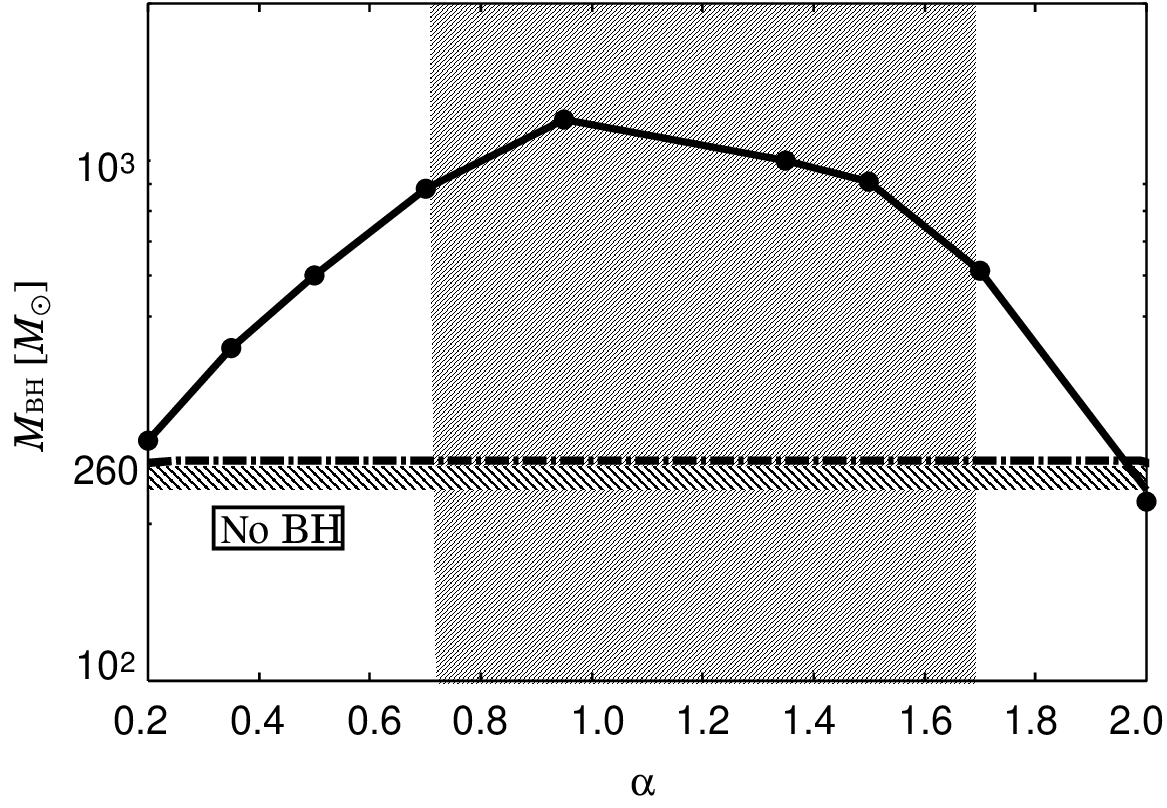}
\figcaption
{
The dependence of the final BH mass ($M_{\rm BH}$ in units of $M_{\odot}$) 
on the IMF slope ($\alpha$). Here, $M_{0}=10^{8} M_{\odot}$
and $k=8.6 {\rm Gyr}^{-1}$ are assumed.
The solid line shows our prediction.
The shaded region satisfies to the observed mass-to-luminosity
ratio and metallicity in globular clusters.
The BH mass is maximal around $\alpha \simeq 1$.
}

\subsection{ Dependence on Star Formation Rate}

In this section, to examine the dependence of
the BH mass on the star formation history in globular clusters,
we alter the coefficient $k$ from $0.1{\rm Gy}^{-1}$ 
to $300{\rm Gy}^{-1}$. 
Here, we assume $M_{0}=10^{8}M_{\odot}$, and set $\alpha=0.95$
which gives the maximal radiation drag efficiency for the BH formation,
as shown in the previous section. 
In Figure 3, the final BH mass is shown against $k$. The corresponding
star formation timescale, which is defined by $t_{\rm SF}= k^{-1}$,
is also shown on the upper horizontal axis. 
The maximal optical depth at ${\it U}$-band is attached
on each point simulated. 
The shaded region satisfies
the observed mass-to-luminosity ratio and
metallicity of globular clusters.
As seen in Figure 3, 
if $t_{\rm SF}$ is shorter than $\sim 10^{8}{\rm yr}$,
the BH mass becomes lower. This is because
$t_{\rm SF}$ is shorter than
the typical age of intermediate mass stars,
$t_{*}\approx 10^{8}{\rm yr}$,
the optical depth cannot increase to a high level before $t_{\rm w2}$.
In contrast, if $t_{\rm SF}$ is longer than $t_{*}$, 
massive BHs with $\ga 10^{3}M_{\odot}$ can form, 
because the optical depth of the system can
reach a high level.
However, if $t_{\rm SF}$ is $\approx 10^{10}$yr,
the number of intermediate mass stars increase slowly with time.
Hence, the BH cannot grow rapidly. Hence,
the BH mass is almost saturated for longer $t_{\rm SF}$.
As a result, we find that, as long as the mass-to-luminosity ratio 
and metallicity satisfy those observed, the final BH mass
is almost regardless of the star formation rate, and
massive BHs with $\approx 10^{3}M_{\odot}$ are
expected to form.

\vspace{5mm}
\epsfxsize=8cm 
\epsfbox{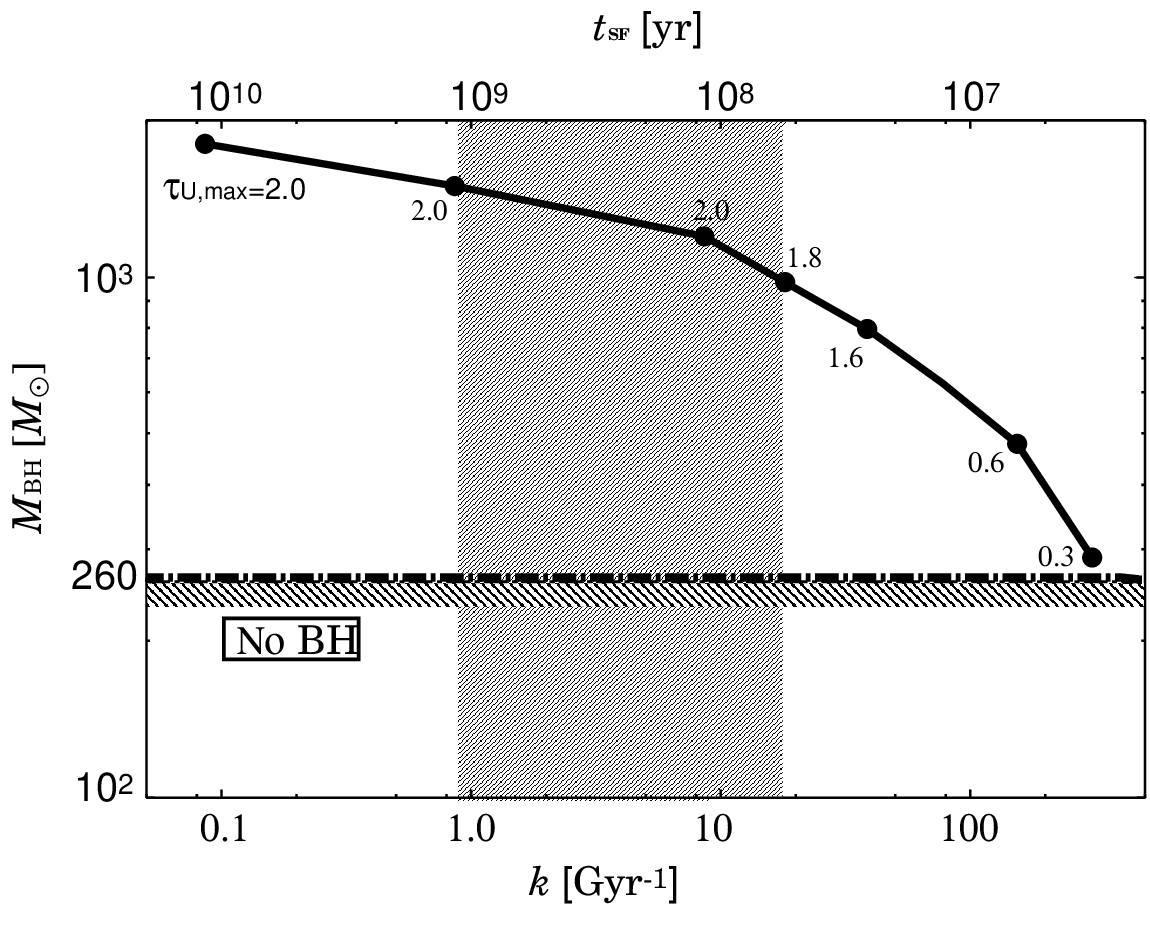}
\figcaption
{
The final BH mass ($M_{\rm BH}$) 
against the star formation rate coefficient ($k$),
assuming $M_{0}=10^{8}M_{\odot}$ and $\alpha=0.95$.
The abscissa is the coefficient in units of Gyr$^{-1}$, 
and the ordinate is the final BH mass in units of solar mass.
The upper horizontal axis is the star formation timescale, $t_{\rm SF}$ 
in units of yr.
The attached values are the maximal optical depth at ${\it U}$-band.
For $t_{\rm SF} > 10^{8}$yr, the BH mass is almost saturated.
For $t_{\rm SF} < 10^{8}$yr, the BH mass decrease rapidly, because
the system becomes optically thin.
The shaded region satisfies the observed mass-to-luminosity ratio 
and metallicity in globular clusters 
($Z_{*}=0.01-0.1Z_{\odot}$).
}

\subsection{ Globular Cluster-to-BH Mass Relation}

Finally, we derive the relation between the final BH 
mass, $M_{\rm BH}$, and the final stellar mass of globular clusters,
$M_{\rm star,final}$.
Here, we set $k$ to be constant as $k=8.6 {\rm Gyr}^{-1}$C
but $\alpha$ is changed.
We consider the range of initial mass as $M_{0}=10^{6-9}M_{\odot}$,
because we focus on the final mass range of 
$M_{\rm star, final}=10^{5-8}M_{\odot}$.
The results are shown in Figure 4.
As shown in previous sections, the final BH mass 
is almost independent of the stellar initial
mass function and star formation rate,
as long as the resultant properties of globular clusters are
consistent with the observations.
In Figure 4, the shaded area is the region which satisfies the
observed mass-to-luminosity ratio and metallicity of globular clusters.
The region of no BH means that
the mass of MDO is less than $260M_{\odot}$, which is the
present criterion for a massive BH. 
Figure 4 shows that massive BHs can form only
for the final mass of globular clusters with $\ga 6\times 10^{6}M_{\odot}$.
The putative BH-to-bulge mass relation, $M_{\rm BH}=0.001M_{\rm bulge}$, 
is shown by a dashed line.
We can see that the predicted BH-to-globular 
cluster mass ratio ($M_{\rm BH}/M_{\rm star, final}$)
is considerably lower than the BH-to-bulge mass ratio.
The main reason is that the duration when the system 
is optically thick is shorter in globular clusters than that in galactic bulges,
because the supernova explosions are more devastating
for smaller spheroidal systems like globular clusters.
It is also noted that the relation between the BH mass and 
the globular cluster mass is not linear, in contrast to
the BH-to-bulge mass relation. 

\vspace{5mm}
\epsfxsize=8cm 
\epsfbox{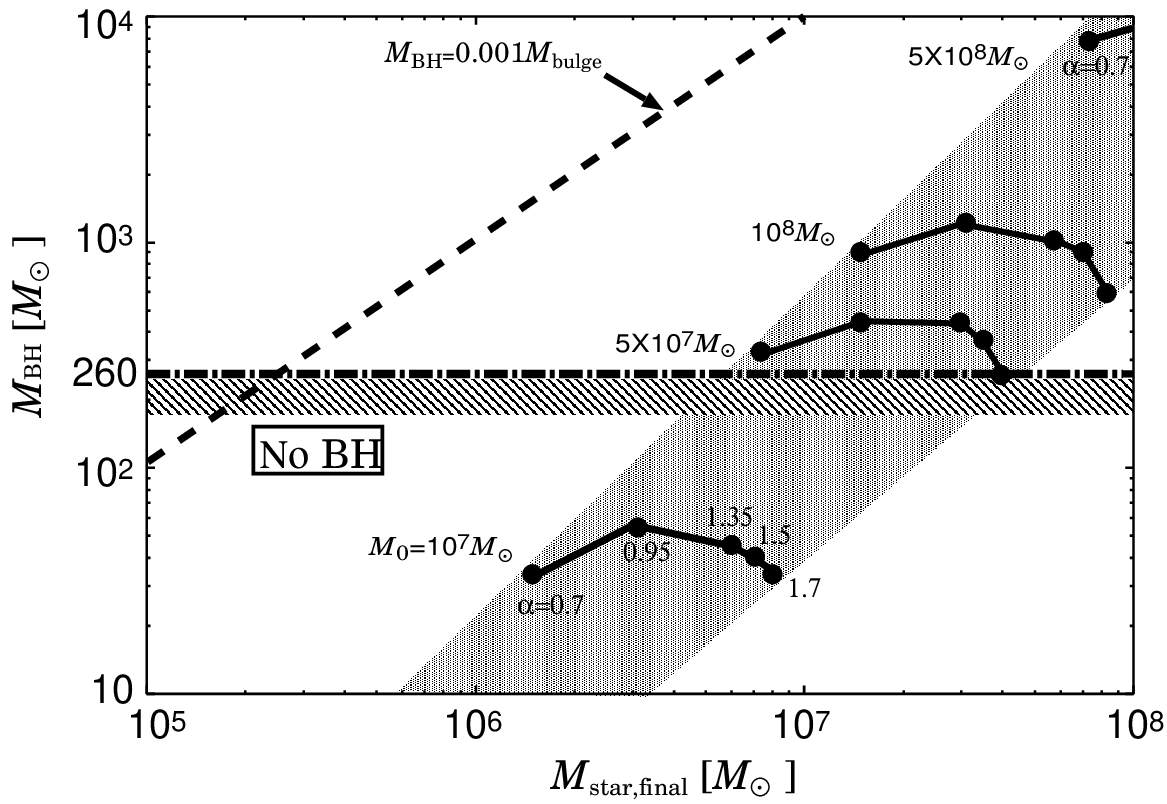}
\figcaption
{
The relation between the final BH mass ($M_{\rm BH}$) and 
the final stellar mass of globular cluster ($M_{\rm star, final}$).
Here, we assume $k=8.6 {\rm Gyr}^{-1}$.
The ordinate is the BH mass in units of solar mass, and 
the abscissa is the final mass of globular cluster in units of solar mass.
The thick solid lines show the results for 
different initial masses of the system, $M_{0}$.
The attached values denote the IMF index $\alpha$.
The shaded area shows the region
which satisfies the observed mass-to-luminosity ratio and metallicity.
The BH-bulge mass relation, $M_{\rm BH}=0.001M_{\rm bulge}$, 
is shown by a dashed line.
The predicted BH fraction is considerably smaller than $10^{-3}$.
}

\section{Comparison with Observations}

To compare the present prediction with the observations
on the massive BHs in globular clusters,
we translate the obtained $M_{\rm BH}-M_{\rm star,final}$ 
relation into $M_{\rm BH}-\sigma$ relation, based
on the virial theorem, where $\sigma$ 
is the stellar velocity dispersion in globular clusters.
The resultant $M_{\rm BH}-\sigma$ relation is shown in Figure 5.
Here, we compare the prediction with BH candidates in M15, G1, 
$\omega$ Cen, the M33 nucleus, and the compact X-ray sources in M82.

In a globular cluster M15 in the Milky Way, the BH mass is estimated to be
$M_{\rm BH} =(1.7^{+2.7}_{-1.7}) \times 10^{3}M_{\odot}$
by Gerssen et al. (2003), which includes the possibility of no BH. 
By the detailed comparison with numerical simulations, 
Baumgardt et al. (2003a) claim that a massive BH with $> 500-1000M_{\odot}$
is not indispensable to account for the observational data in M15.
The total stellar mass of M15 is roughly $10^{6}M_{\odot}$ 
and the velocity dispersion is $\sigma \approx 14{\rm km/s}$.
As seen in Figure 5, the present model predicts that no massive 
BH forms for the velocity dispersion in M15. Hence, the prediction
of the present model supports the case of no BH. 

In a globular cluster G1 in M31, a candidate for a massive BH
with $M_{\rm BH}\approx 2\times 10^{4}M_{\odot}$ is reported
(Gebhardt, Rich \& Ho 2002). But, Baumgardt et al. (2003b) argue
that the observational data in G1 can be well fitted even
without a massive BH. Hence, the above mass of BH in G1 
should be considered as an upper limit. 
The total mass of G1 is $ (0.7-2) \times 10^{7}M_{\odot}$ and
the velocity dispersion is estimated as 
$\sigma \approx 25{\rm km/s}$ (Meylan et al. 2001).
It is predicted for this velocity dispersion 
that no massive BH forms if the system mass is initially lower 
than $5 \times 10^{7}M_{\odot}$, while 
a BH with $\approx 500M_{\odot}$ can form
if the system mass is initially higher than 
that mass. 

In addition, $\omega$ Cen possibly harbors a massive BH.
$\omega$ Cen is a globular cluster in the Milky Way 
in which member stars show a wide range of metallicity.
The total stellar mass of $\omega$ Cen 
is $5\times 10^{6}M_{\odot}$, and the
velocity dispersion is $\sigma \approx 22{\rm km/s}$ (Meylan et al. 1995).
As seen in Figure 5, whether $\omega$ Cen can 
have a massive BH with $\approx 300M_{\odot}$ is marginal.
Latest radio observations have suggested that 
the black hole mass in $\omega$ Cen should be less than about 
$100 M_{\odot}$ for the spherical Bondi-Hoyle accretion rate, but
the data are marginally consistent with a black hole of 
about $1000 M_{\odot}$ for more plausible accretion rate
 (Maccarone et al. 2005).

\vspace{5mm}
\epsfxsize=8cm 
\epsfbox{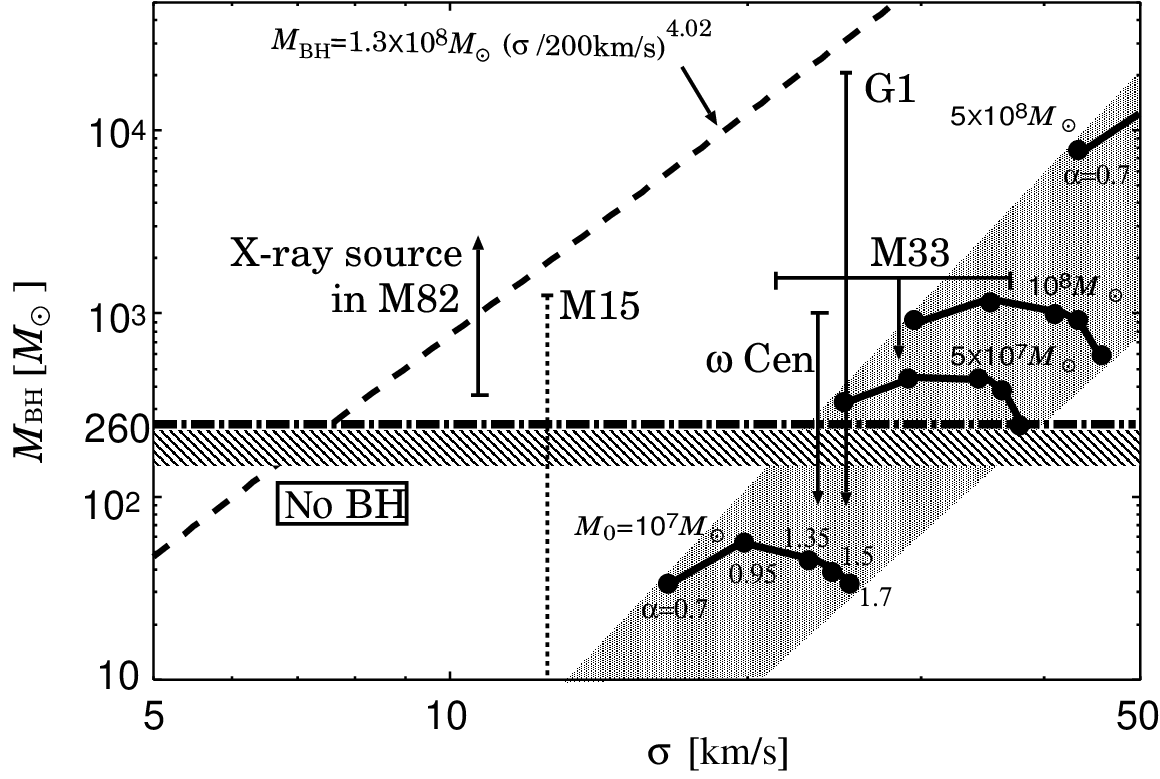}
\figcaption
{
Comparison of the theoretical prediction with the observational data 
on BH mass in globular clusters.
The ordinate is the BH mass ($M_{\rm BH}$) in units of solar mass, 
and the abscissa is the velocity dispersion of globular cluster ($\sigma$) 
in units of km/s. The parameters are the same as Figure 4.
The shaded region is the prediction of the present analysis. 
The relation between the BH mass and the velocity dispersion of galactic 
bulges is shown by a dotted line, $M_{\rm BH}=1.3\times 10^{8}M_{\odot}
(\sigma/200{\rm km}/{\rm s})^{4.02}$ (Tremaine et al. 2002).
The observational data of M15, G1, $\omega$ Cen, M33 nucleus and M82 are plotted.
}
\vspace{2mm}

On the other hand, no strong evidence of a massive BH
is found in a stellar cluster at the center of M33
 (Merritt, Ferrarese, \& Joseph 2001; Gebhardt et al. 2001).
The M33 nucleus has the velocity dispersion of
$\sigma =21-34{\rm km/s}$ (Merritt, Ferrarese, \&
Joseph 2001). For this velocity dispersion, our model
predicts that a massive BH with $\approx 10^{3}M_{\odot}$
can form only if the initial system mass is higher than 
$5 \times 10^{7}M_{\odot}$.

Finally, Matsumoto et al. (2001) have identified 
nine bright compact X-ray sources in M82 by Chandra observatory.
The brightest source (No.7 in their Table 1) has
a luminosity of $9\times 10^{40}{\rm erg}\, {\rm s}^{-1}$.
If this luminosity is assumed to be the Eddington luminosity,
the derived mass of a black hole is $350M_{\odot}$.
From the infrared luminosity, 
the total mass of the cluster is estimated to be 
$\sim 5\times 10^{5}M_{\odot}$ and the velocity dispersion is 
to be $11{\rm km/s}$ (McCrady et al. 2003). 
As seen in Figure 5, the present model predicts
that no BH forms for this velocity dispersion,
although the possibility of BH formation via
runaway stellar collision is proposed (Portegies Zwart \& McMillan 2002; Portegies Zwart et al. 2004).

\section{Discussion} 
In the previous sections (\S 3 and \S4), we have shown that massive 
BHs are expected to be hosted by globular clusters. 
Here, we consider the detectability of the BH formation in globular clusters.
The present scenario of BH formation is summarized as follows: 
(i) First, a massive self-gravitating viscous disk (MDO) forms 
owing to the radiation drag in a timescale of $\approx 10^{8}-10^{9}{\rm yr}$,
and the core evolves into a VMS. 
(ii) Secondly, the VMS collapses directly into a BH if its mass 
is higher than $260M_{\odot}$. 
(iii) Finally, the BH grows to a massive BH by the mass accretion from 
the ambient gas. 

A VMS more massive than $260M_{\odot}$ evolves in a short 
lifetime of $\le 10^{6}{\rm yr}$ (e.g., Schaerer 2002).
Therefore, the stages (i) and (ii) correspond to a high redshift, 
if we consider the age of globular clusters. Hence, it would be hard 
to detect a VMS itself, because the apparent magnitude is too faint to
observe.
However, a VMS may result in a gamma-ray burst (GRB), if 
it is Population II and a forming BH 
entails the sufficient angular momentum (Heger et al. 2003). 
GRBs are detectable even at redshift higher than 10 
(Lamb \& Reichart 2001). 
The present scenario predicts that GRB events would 
take place at the centers of forming globular clusters 
at early stages of $\approx 10^{8}-10^{9}{\rm yr}$. 

In the stage (iii), the luminosity of BH accretion
is expected to be $L=\eta \dot{M} c^2$, 
where $\dot{M}$ is the mass accretion rate and $\eta$ is the efficiency.
The temperature in the innermost regions of accretion disk
and the efficiency $\eta$ depend on whether $\dot{M}$ is higher or lower
than the Eddington accretion rate $\dot{M}_{\rm E}=L_{\rm E}/c^2$
(e.g. Kato, Mineshige, \& Fukue, 1998), where $L_{\rm E}$ is the 
Eddington luminosity given by
\begin{equation}
L_{\rm E}=1.25\times 10^{41}\,{\rm erg}\, {\rm s}^{-1}
\left(\frac{M_{\rm BH}}{10^{3}M_{\odot}}\right).
\end{equation}
If $\dot{M} \approx \dot{M}_{\rm E}$, a so-called {\it standard disk} forms.
Then, the efficiency is $\eta \approx 0.1$ and
the temperature is given by 
\begin{equation}
T_{\rm disk}=5.8\times 10^{6}\, {\rm K}\, 
\left(\frac{M_{\rm BH}}{10^{3}M_{\odot}}\right)^{-1/4}
\left(\frac{r}{3r_{\rm sch}}\right)^{-3/8},
\end{equation}
where $r_{\rm sch}$ is the Schwarzschild radius.
If $\dot{M} \ll \dot{M}_{\rm E}$, the solution of accretion is optically
thin and called a radiatively inefficient accretion flow ({\it RIAF}). 
Then, $\eta \approx 0.1(\dot{M}/\dot{M}_{\rm E})$ and 
the temperature is on the order of $10^9$ K for electrons.
If the accretion rate is super-Eddington ($\dot{M} > \dot{M}_{\rm E}$), 
the solution is optically-thick and called a {\it slim disk}, where
the temperature is higher than the standard disk.
In this type of disk, 
the photon-trapping effect in the accretion flow plays an important role
and then the efficiency becomes 
$\eta \approx 0.1(\dot{M}/\dot{M}_{\rm E})^{-1/2}$.
Actually, the accretion luminosity can achieve up to
$\approx 7L_{\rm E}$ (Ohsuga et al. 2002, 2003).

Thus, if $\dot{M} \ga \dot{M}_{\rm E}$, it is expected that 
an accreting BH in a globular cluster would be a far-UV and X-ray source
with $L \ga L_{\rm E}$. If a slim disk is considered,
$L \approx 10^{41}{\rm erg}\,{\rm s}^{-1}$ is realized even for
a BH with several $100M_{\odot}$. 
Such a source may be observed as a ultraluminous X-ray source (ULX), 
the luminosity of which is $10^{39}-10^{41}{\rm erg}\,{\rm s}^{-1}$.
The lifetime of the accretion phase is pretty shorter than the present age of
globular clusters. But, in some globular clusters, the final accretion phase
may be detected. 
In fact, Liu \& Bregman (2005) have recently reported that 
some ULXs are located at old globular clusters.
These ULXs can be the candidates for accreting BHs 
with $10^{3}M_{\odot}$ formed by the radiation hydrodynamic process.

\section{Conclusions}
Based on the radiation hydrodynamic model,
we have examined the possibility for the formation 
of massive BHs in globular clusters, with
incorporating the realistic chemical evolution.
In the present model, the mass of a central massive 
dark object in a globular cluster is determined by the
duration of the optically-thick phase.
We have found that massive BHs can form only
in globular clusters with the the final mass of
$\ga 6\times 10^{6}M_{\odot}$
or the velocity dispersion of $\ga 20{\rm km/s}$.
The BH mass 
is almost independent of the stellar initial
mass function and star formation rate.
The results show that massive BHs are not likely to form
in typical globular clusters with $\approx 10^6M_{\odot}$.
The predicted BH-to-globular 
cluster mass ratio ($M_{\rm BH}/M_{\rm star, final}$)
is considerably lower than the BH-to-bulge mass ratio.

We have applied the present model to some globular clusters, 
in which the possibilities of massive BHs are reported.
In M15, no massive BH is predicted to form,
although the evidence of a massive BH is under debate.
G1 can harbor a massive BH with $\approx 500M_{\odot}$,
which is smaller by an order of magnitude than
an estimate, $M_{\rm BH}\approx 2\times 10^{4}M_{\odot}$,
by Gebhardt, Rich \& Ho (2002).
In $\omega$ Cen, the formation of a massive BH with 
$\approx 300M_{\odot}$ is marginal. 
As for M82, it is suggested that massive BHs with $350M_{\odot}$
can reside in bright compact X-ray sources. But, 
no BH is expected to form in the present radiation hydrodynamic model. 
Finally, the present scenario predicts that a final phase of BH growth 
in a globular cluster can be observed as a ULX with 
$\sim 10^{41}{\rm erg}\,{\rm s}^{-1}$. 
Also, GRBs may arise in early phases ($10^{8}-10^{9}{\rm yr}$)
of globular cluster formation.

In this paper, we have investigated the possibility of the BH formation
from a hydrodynamic point of view.
There may be a different path to form massive BHs in globular clusters.
To eclucidate the mechanism for the formation of massive BHs, 
the high resolution observations due to the multi-wavelength, especially optical, far-UV, X-ray and gamma-ray on the central regions of globular clusters would be greatly important in the future.

\acknowledgments
The authors thank the anonymous referee for his/her fruitful 
comments and suggestions. 
We thank T. Nakamoto and A. Yonehara for many useful comments. 
NK also acknowledges Italian MIUR and INAF financial support. 
Numerical simulations were performed 
with facilities at Center of Computational Sciences,
University of Tsukuba. This work was supported in part 
by Grants-in-Aid for Scientific Research from MEXT, 16002003.


\begin{table}[t]
\begin{center}
Table 1. BH mass-to-IMF Slope in Globular Cluster\\[3mm]
{\scriptsize
\begin{tabular}{cccccc}
\hline \hline
$\alpha$ & $M_{\rm BH}(M_{\odot})$ & $M_{\rm
star,final}(M_{\odot})({\rm a})$ & $\tau_{\rm U,max}({\rm b})$
& $t_{\rm w2}({\rm yr})$ & $(M/L)_{\odot}$ \\ \hline
0.2  & 290 & $3.4\times 10^{5}$ & 1.3 & $9\times 10^{7}$ & 0.1 \\
0.35 & 436 & $3.0\times 10^{6}$ & 1.4 & $10^{8}$ & 0.47 \\
0.5  & 600 & $6.9\times 10^{6}$ & 1.5 & $1.1\times 10^{8}$ & 0.7 \\
0.7  & 880 & $1.5\times 10^{7}$ & 1.5 & $1.2\times 10^{8}$ & 1.0 \\
0.95 & 1200 & $3.0\times 10^{7}$ & 2.0 & $1.4\times 10^{8}$ & 1.3 \\
1.35 & 1000 & $6.0\times 10^{7}$ & 1.2 & $2.0\times 10^{8}$ & 2.0 \\
1.5  & 909  & $7.0\times 10^{7}$ & 0.9 & $2.5\times 10^{8}$ & 2.3 \\
1.7  & 613  & $8.2\times 10^{7}$ & 0.4 & $4.5\times 10^{8}$ & 2.9 \\
2.0  & (221)  & $9.2\times 10^{7}$ & 0.2 & $1.6\times 10^{9}$ & 4.0 \\ \hline
\end{tabular}
}
\noindent
\end{center}
{\scriptsize 
(a) $M_{\rm star, final}$ is the mass of globular cluster in the present day.
(b) $\tau_{\rm U,max}$ represents the maximal optical
depth at {\it U}-band.
}
\end{table}

\end{document}